\documentclass[9pt, conference]{IEEEtran}
\IEEEoverridecommandlockouts

\usepackage{cite}
\usepackage{amsmath,amssymb,amsfonts}
\usepackage{algorithmic}
\usepackage{graphicx}
\usepackage{textcomp}
\usepackage{xcolor}
\def\BibTeX{{\rm B\kern-.05em{\sc i\kern-.025em b}\kern-.08em
    T\kern-.1667em\lower.7ex\hbox{E}\kern-.125emX}}
\begin{document}

\title{
Optimized Self-supervised Training with BEST-RQ \\for Speech Recognition
}

\author{\IEEEauthorblockN{Ilja Baumann, Dominik Wagner, Korbinian Riedhammer, Tobias Bocklet}
\IEEEauthorblockA{
\textit{Technische Hochschule Nürnberg}\\
Nuremberg, Germany \\
\{firstname.lastname\}@th-nuernberg.de}
}


\makeatletter
\def\ps@IEEEtitlepagestyle{%
  \def\@oddfoot{\mycopyrightnotice}%
  \def\@evenfoot{}%
}
\def\mycopyrightnotice{%
  \begin{minipage}{\textwidth}
  \centering \scriptsize
  Copyright~\copyright~2025 IEEE. Personal use of this material is permitted. Permission from IEEE must be obtained for all other uses, in any current or future media, including reprinting/republishing this material for advertising or promotional purposes, creating new collective works, for resale or redistribution to servers or lists, or reuse of any copyrighted component of this work in other works.
  \end{minipage}
}
\makeatother

\maketitle


\begin{abstract}
Self-supervised learning has been successfully used for various speech related tasks, including automatic speech recognition. BERT-based Speech pre-Training with Random-projection
Quantizer (BEST-RQ) has achieved state-of-the-art results in speech recognition. In this work, we further optimize the BEST-RQ approach using Kullback-Leibler divergence as an additional regularizing loss and multi-codebook extension per cluster derived from low-level feature clustering.
Preliminary experiments on train-100 split of LibriSpeech result in a relative improvement of 11.2\% on test-clean by using multiple codebooks, utilizing a combination of cross-entropy and Kullback-Leibler divergence further reduces the word error rate by 4.5\%. 
The proposed optimizations on full LibriSpeech pre-training and fine-tuning result in relative word error rate improvements of up to 23.8\% on test-clean and 30.6\% on test-other using 6 codebooks.
Furthermore, the proposed setup leads to faster convergence in pre-training and fine-tuning and additionally stabilizes the pre-training.
\end{abstract}

\begin{IEEEkeywords}
self-supervised learning, speech recognition, conformer, BEST-RQ
\end{IEEEkeywords}

\section{Introduction}
Self-supervised learning in audio and speech processing has emerged as an effective method for learning useful representations from unlabeled audio data. 
The resulting models leverage unlabeled audio to extract meaningful features without the need for large amounts of labeled data. 
By predicting parts of the audio from other parts or using contrastive learning \cite{hadsell2006dimred} approaches, self-supervised models can effectively capture the temporal and spectral patterns in speech and sound.
This capability makes them highly valuable for various downstream tasks, such as speech recognition, speaker identification, and audio classification, reducing the demands on annotated datasets and enhancing model performance in diverse acoustic environments.

Recent self-supervised speech processing models have been inspired by BERT \cite{devlin2019bert}. In BERT, a percentage of the tokens in the input sequence is randomly masked. 
The goal of the model is to predict the original, masked tokens based on the surrounding context.
Approaches like wav2vec 2.0 \cite{baevski2020wav2vec2} employ a contrastive learning objective where speech features are quantized into a set of target feature vectors and the positive and negative features are used as target features.
Wav2vec 2.0 focuses on contrastive learning by masking and predicting raw audio signal representations, whereas HuBERT \cite{hsu2021hubert} uses pseudo-labels generated by clustering to predict hidden units of speech during the masking process.
Target tokens are generated in the first step on frame-level by using k-means clustering on MFCCs. 
In the second step, targets are generated by using intermediate embeddings of the trained model from step one, also employing k-means for clustering.

To further improve the performance on downstream tasks, especially the paralinguistic and speaker-id related downstream tasks, WavLM \cite{chen2022wavlm} was proposed. A noisy speech augmentation technique involves adding a segment of speech from a different speaker to the current speech, paired with a denoising masked token prediction objective. The model is trained to predict the target tokens derived from the original clean speech.

In w2v-BERT \cite{Chung2021w2vBERT}, a combination of both objectives of wav2vec 2.0 and HuBERT was proposed. Contrastive loss is applied on the middle layer outputs and predictive loss like in HuBERT at the final layer output.

Subsequently, a BERT-based Speech pre-Training with Random-projection Quantizer (BEST-RQ) \cite{chiu2022bestrq} setup for self-supervised learning demonstrated that the clustering-based token generation from HuBERT could be effectively replaced with a fixed random-projection quantization approach. 
This simple modification was shown to match or even surpass the performance of HuBERT on automatic speech recognition (ASR) tasks.

Codebooks for specific speech and speaker characteristics have been examined in \cite{prabhu2023accented}, where codebooks per accent were used to capture accent-specific information.

In this study, we further enhance the BEST-RQ training setup for self-supervised learning and test our approach on the downstream task of speech recognition.
Our contributions are:
\begin{itemize}
    \item The analysis on the effect of the codebook size.
    \item The analysis on the effect of multiple codebooks in pre-training.
    \item Optimized pre-training by employing Kullback-Leibler divergence loss for regularization.
    \item Cluster-specific codebooks derived from low-level audio features.
\end{itemize}

\begin{figure*}[htbp]
\centerline{\includegraphics[width=0.86\linewidth]{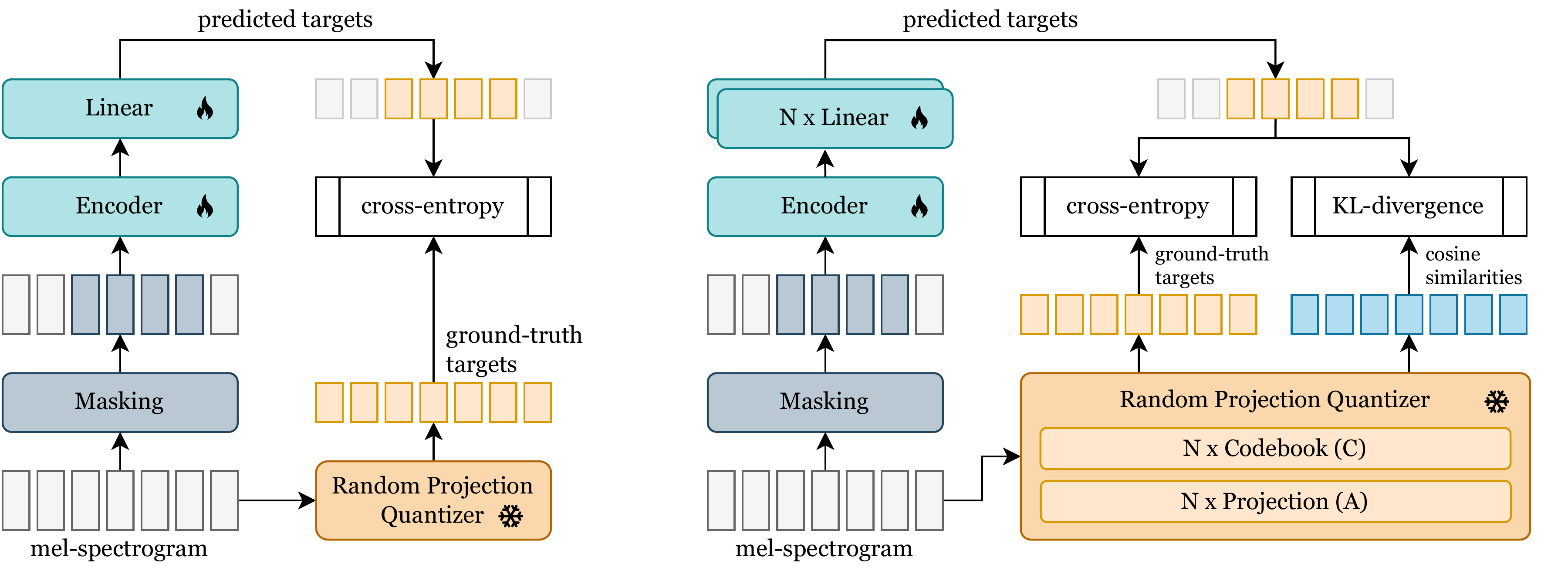}}
\caption{BEST-RQ training setup: (left) shows the baseline setup of BEST-RQ proposed in \cite{chiu2022bestrq}, (right) shows our proposed modifications, including KL-divergence as regularizing loss and multiple codebooks in pre-training.}
\label{fig:overview}
\end{figure*}

\section{Background}
\label{sec:background}

\subsection{BEST-RQ}
BERT-based Speech pre-Training with Random-projection Quantizer (BEST-RQ) is a self-supervised learning approach for ASR.
The primary idea is to use a random-projection quantizer (RQ) to map speech signals into discrete labels, which are subsequently used as targets during training. 
The RQ projects input features by employing a fixed projection matrix to a vector matching the codebook dimension. 
This vector is used to find the most similar codebook entry, the index of the codebook entry is the target. 
This approach simplifies the model architecture and prevents codebook collapse during training.
The training setup is illustrated in Figure \ref{fig:overview} (left).

\subsubsection{Masking}
The input speech data are 80-dimensional log-mel filter bank coefficients, which are first normalized to have a mean of 0 and a standard deviation of 1. 
This normalization prevents the random projections from collapsing to a limited subset of the codebook. 
The input is randomly masked: at each frame, with a fixed probability of 15\% that a mask is applied, spanning from a starting frame over a length of 4 frames. 
The masked sections are replaced with Gaussian noise with a mean of 0 and a standard deviation of 0.1.

\subsubsection{Random-Projection Quantizer}
The random-projection quantizer is initialized with a randomly generated matrix $A$ and a codebook $C$. 
Input speech signals are projected using a matrix $A$, and each projection is matched to the nearest vector in the codebook $C$. 
This vector's index serves as the label. 
Both, the projection matrix and the codebook remain fixed during training, eliminating the need for updating the quantizer parameters.
This fixed nature of the quantizer ensures that the model's training focus is solely on learning to predict the masked portions of the input based on the unmasked ones.

\subsubsection{Pre-Training}
The pre-training phase involves adding a layer on top of the encoder, which is a stack of conformer \cite{gulati2020conformer} layers in the original setup \cite{chiu2022bestrq}. 
The model's objective is to predict the quantized labels for the masked sections of the speech input, employing cross-entropy. 
This configuration provides flexibility, as the random-projection quantizer is independent of the encoder.

\subsubsection{Fine-Tuning}
A supervised dataset is used to fine-tune the model after pre-training. 
The encoder, initialized with the pre-trained model parameters, is trained for downstream tasks. 
During this phase, a projection layer is added on top of the pre-trained encoder to fine-tune for the downstream speech recognition task.

\section{Methods}
\label{sec:methods}
To optimize the BEST-RQ pre-training we conduct several optimizations: multi-codebook setup, auxiliary loss for regularization and cluster specific codebooks derived from acoustic features.

\subsection{Multi-Codebook}
We follow \cite{Zhang2023GoogleUSM} to improve BEST-RQ training by using multiple codebooks rather than a single codebook.
Specifically, we employ $N$ output layers to generate $N$ probability predictions from the encoder's output, which are then compared to $N$ independent quantization targets derived from the masked speech features. 
Each layer loss is given equal weight during training.

\subsection{Feature Cluster-Specific Codebooks}
\label{subsec:feature-cluster-spec-codebooks}
Instead of only using equally weighted codebooks, we cluster the audio files utterance-wise on feature-level.
For this, we compute the following features for each utterance and take the mean vectors for clustering:

\begin{itemize}
    \item MFCC (Mel-Frequency Cepstral Coefficients) \cite{davis1980repr}: MFCCs are representations of the short-term power spectrum of a sound, using a mel scale, and capture essential information about the speech signal's timbre and pitch.
    \item Spectral Contrast \cite{jiang2002spectralcontrast}: Measures the difference in amplitude between peaks and valleys in a sound spectrum, capturing the spectral distribution's shape. It is beneficial as it highlights the tonal and harmonic content, which helps distinguish between different types of sounds or speech qualities.
    \item Spectral Roll-Off: This parameter indicates the frequency below which a specified percentage (set to 85\% in this work) of the total spectral energy lies, effectively summarizing the spectral content's distribution. It can help in clustering by identifying sounds with similar energy distributions, which often correspond to similar types of speech or acoustic events.
    \item Zero-Crossing Rate: Counts how often the signal waveform crosses the zero amplitude axis, which is related to the signal's noisiness and texture. It is useful for differentiation between voiced and unvoiced sounds, providing insights into the rhythm and articulation of speech.
\end{itemize}

Each utterance is assigned to a particular cluster. 
In pre-training, each cluster is assigned to a particular codebook, e.g., using 6 feature clusters would result in 6 codebooks in pre-training.
In pre-training, the loss is weighted with a primary weight $w_p$ for each sample belonging to a particular cluster, the remaining codebook losses with a secondary weight $w_s$. 
We choose $w_p$ to be at least twice as high as $w_s$.

\subsection{Auxiliary Loss}
\label{subsec:kl-loss}
Beside using cross-entropy loss between the predicted targets and ground truth indices, we employ KL-divergence as an auxiliary loss for regularization that helps the model better generalize to unseen data. 
KL-divergence, in this context, measures the difference between the predicted distribution and the target distribution, ensuring that the model's output is not only accurate but also consistent with the expected probabilistic structure of the data.
By minimizing this divergence, the model learns to align its predictions more closely with the actual distributions of the target features across different codebooks.
This additional regularization mitigates overfitting by encouraging smoother and more robust predictions, particularly in complex datasets where variations in speech patterns and noise can introduce instability.
Furthermore, using KL-divergence helps reduce the variance between predictions across different parts of the input space, leading to more stable training.
This stability enables faster convergence, as the model avoids sharp shifts in parameter updates that can result from over-reliance on cross-entropy loss alone.
The combination of cross-entropy and KL-divergence leads to improved generalization, making the model more effective in real-world speech recognition tasks where diverse and noisy data are common.

Let $\mathbf{p}^{(n)}$ denote the prediction output from the $n$-th codebook, and let $\mathbf{t}^{(n)}$ represent the target output corresponding to that codebook. 
Given $N$ codebooks, the prediction outputs can be denoted as $\{\mathbf{p}^1, \mathbf{p}^2, \ldots, \mathbf{p}^N\}$ and the targets as $\{\mathbf{t}^1, \mathbf{t}^2, \ldots, \mathbf{t}^N\}$.
The overall loss function $L$ is a weighted combination of the Cross-Entropy (CE) loss and the Kullback-Leibler (KL) divergence loss.

Cross-Entropy loss is computed as:
\begin{align}
L_{\text{CE}} = \frac{1}{N} \sum_{n=1}^{N} \left( - \sum_{i=1}^{M} \mathbf{t}_{i}^{(n)} \log\left(\mathbf{p}_{i}^{(n)}\right) \right)
\end{align}

where the standard cross-entropy loss between the predictions $\mathbf{p}_i$ and the targets $\mathbf{t}_i$ is calculated over all $N$ codebooks. $M$ represents the number of targets.

For each codebook's prediction and similarity distribution, we independently compute KL divergence loss as:


\begin{align}
L_{\text{KL}} &= \frac{1}{N} \sum_{n=1}^{N} \sum_{i=1}^{M} \mathbf{p}_{i}^{(n)} \log\left(\frac{\mathbf{p}_{i}^{(n)}}{\mathbf{d}_{i}^{(n)}}\right)
\end{align}

where $\mathbf{p}_i$ and $\mathbf{d}_i$ are the codebook-specific predictions and similarity vectors, respectively. Each KL divergence term is subsequently scaled by a weight, $w_{\text{KL}}$ resulting in an overall objective:

\begin{align}
L = w_{\text{CE}} \cdot L_{\text{CE}} + w_{\text{KL}} \cdot L_{\text{KL}}
\end{align}

We use $w_{\text{KL}} = 0.1$ and $w_{\text{CE}} = 1$ as scaling factors, which results in approximately equally weighted losses as a result of empirical tests.

Figure \ref{fig:overview} shows the baseline setup (left) compared to our proposed modifications (right) in pre-training.

\section{Experiments}
We use BEST-RQ with conformer \cite{gulati2020conformer} encoder as our baseline, as described in Section \ref{sec:background}. 
The speech encoder consists of a stack of 12 conformer layers, following the setup in \cite{whetten2024bestrq}.
Each layer consists of multi-head self-attention comprised of 8 heads.
Our baseline is the implementation in \cite{whetten2024bestrq}, the experiments are conducted using SpeechBrain \cite{speechbrain, ravanelli2024opensourceconversationalaispeechbrain}.
Instead of using L2 to select the target index, we employ cosine similarity, as proposed in \cite{Zhang2023GoogleUSM}.

We use LibriSpeech \cite{panayotov2015librispeech} for all our experiments. 
The raw input speech is converted to 80-dimensional log-mel filter bank coefficients and used as inputs.
We conduct preliminary experiments, exploring our modifications using only the train-100 split in training for a small amount of training steps.
The models are trained for 25 epochs, approximately 33k steps, on a single GPU.
We chose the limited pre-training setup to be able to run our experiments within a limited amount of time using only a limited number of GPU hours, that are available in an academic environment.

\subsection{Preliminary Experiments}
In our preliminary experiments, we fine-tune the pre-trained models using train-100 split and evaluate on test-clean.

For the codebook size experiments, we vary the codebook size and dimensionality of codebook vectors of the single codebook setup.
We examine lengths of $l=\{4096, 8192, 10240\}$ and dimensions of $d=\{16, 32, 64\}$.
For the multi-codebook training, we employ $N$ codebooks
$N \in \{x \mid x \in \mathbb{Z}, 1 \leq x \leq 10, x \text{ is even}\}$.

\subsection{Codebook Size and Dimension}
Results for the codebook size experiments are listed in Table \ref{tab:cb-size}.
Baseline is the setup underlined with a length of 8192 and dimension of 16.
The best performing configuration is with 10240 length and a 32 dimensional codebook.
The relative WER improvement on dev-clean and test-clean is 4.7\% and 4.2\% respectively compared to the baseline.

\begin{table}[htbp]
\caption{WERs fine-tuned using train-100 for different codebook sizes. The \underline{underlined} line represents the baseline setup.}
\begin{center}
\begin{tabular}{|c|c|c|c|}
\hline
\multicolumn{2}{|c|}{\textbf{Codebook}} & \multicolumn{2}{|c|}{\textbf{WER}} \\
\hline
\textbf{size} & \textbf{dimension} & \textbf{dev-clean} & \textbf{test-clean} \\
\hline
4096    & 16  & 18.79   & 19.00 \\
\underline{8192}    & \underline{16}  & \underline{19.21}   & \underline{19.47} \\
10240    & 16  & 19.38   & 19.66 \\
4096    & 32  & 19.67   & 20.10 \\
8192    & 32  & 19.85   & 19.96 \\
10240    & 32  & \textbf{18.31}   & \textbf{18.66} \\
4096    & 64  & 20.25   & 20.47 \\
8192    & 64  & 19.88   & 19.90 \\
10240    & 64  & 19.23   & 19.16 \\
\hline
\end{tabular}
\label{tab:cb-size}
\end{center}
\end{table}

\subsection{Multi-Codebook Training}
The results for the multi-codebook setup as proposed in \cite{Zhang2023GoogleUSM} are listed in Table \ref{tab:multi-codebook-results}.
The performance of all multiple codebooks tested surpasses that of the baseline setup using only one codebook. 
The best WER is achieved by using 6 codebooks, resulting in a relative improvement of 11.2\% on test-clean. Although the WER on dev-clean is the lowest when employing 4 codebooks, the results obtained between 4 and 6 codebooks are not statistically significant. Therefore, we selected 6 codebooks for further experiments.

\begin{table}[htbp]
\caption{WERs fine-tuned using train-100 for different multi-codebook setups. The \underline{underlined} line represents the baseline setup, \textbf{bold} the best result, * the second-best result.}
\begin{center}
\begin{tabular}{|c|c|c|c|}
\hline
\textbf{Codebook} & \multicolumn{2}{|c|}{\textbf{WER}} \\
\hline
\textbf{N} & \textbf{dev-clean} & \textbf{test-clean} \\
\hline
\underline{1} & \underline{19.21}   & \underline{19.47} \\
2  & 18.52   & 18.42 \\
4  & \textbf{17.07}   & 17.40* \\
6  & 17.22*   & \textbf{17.29} \\
8  & 17.46   & 17.78 \\
10  & 17.64   & 17.83 \\
\hline
\end{tabular}
\label{tab:multi-codebook-results}
\end{center}
\end{table}

\subsection{Auxiliary Loss}
We utilize KL-divergence as a regularizing loss as described in Section \ref{subsec:kl-loss}, results are shown in Table \ref{tab:kl-div-results}.
Using only KL-divergence results in slightly worse results, yet this could probably be mitigated with a longer pre-training.
A combination of CE and KL-divergence outperforms only using CE on dev-clean and test-clean by 7.8\% and 8.4\% respectively using one codebook.
Using 6 codebooks and the loss combination, the improvement on dev-clean and test-clean is 4.6\% and 4.5\% respectively, compared to CE loss only and 6 codebooks.
Using KL-divergence and CE with equal weighting compared to CE not only leads to better results, but also faster convergence.
Using our modified pre-training method, we achieve the lowest validation loss of the baseline after a third of the training steps.

\begin{table}[htbp]
\caption{WERs fine-tuned using train-100 for extended loss modifications. The \underline{underlined} line represents the baseline setup, \textbf{bold} the best result, * the second-best result.}
\begin{center}
\begin{tabular}{|c|l|c|c|c|}
\hline
\textbf{Codebook} & \textbf{Loss} & \multicolumn{2}{|c|}{\textbf{WER}} \\
\hline
\textbf{N} & & \textbf{dev-clean} & \textbf{test-clean} \\
\hline
\underline{1} & \underline{CE} & \underline{19.21}   & \underline{19.47} \\
1  & KL-div & 19.71   & 19.85 \\
1  & CE + KL-div& \textbf{17.71}   & \textbf{17.84} \\
\hline
6  & CE & 17.22   & 17.29 \\
6  & KL-div & 18.61   & 18.10 \\
6  & CE + KL-div & \textbf{16.43}   & \textbf{16.51} \\
\hline
\multicolumn{4}{|c|}{feature-level cluster-specific codebooks} \\
\hline
6  & CE + KL-div & \textbf{16.13}   & \textbf{16.28} \\
\hline
\end{tabular}
\label{tab:kl-div-results}
\end{center}
\end{table}

\subsection{Cluster-Specific Codebooks}
We further conduct experiments using clustering on utterance level, as described in Section \ref{subsec:feature-cluster-spec-codebooks}.
For this we use the best performing setup from the preceding experiments with 6 codebooks, and combined loss of CE and KL-divergence.
Results in Table \ref{tab:kl-div-results} show that using simple weighting of codebooks by cluster leads to a slight improvement in WER.
We further observe, that cluster-specific codebooks lead to a more stable training, considering the validation loss.

\subsection{Full Pre-Training and Fine-Tuning}
To test whether our proposed training modifications scale to the full pre-training, we use all train splits of LibriSpeech, resulting in combined 960 hours for pre-training.
We train the base model and proposed modified model for 100k steps using an effective batch size of 1 hour.

\begin{figure}[htbp]
\centerline{\includegraphics[width=0.84\linewidth]{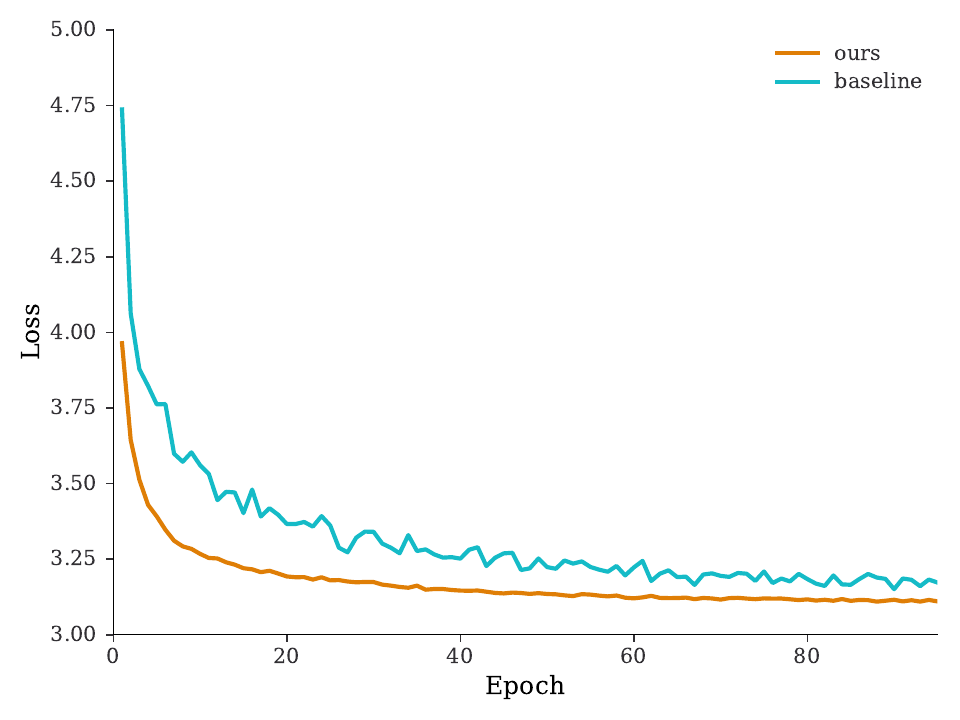}}
\caption{Validation loss in pre-training for the baseline method and our proposed modified configuration.}
\label{fig:pretraining-val-loss}
\end{figure}

Results in Table \ref{tab:full-results} show an improvement over all dev and test sets.
The proposed optimizations result in an improvement of WER of up to 23.8\% on test-clean and 30.6\% on test-other splits.
Furthermore, the validation loss illustrated in Figure \ref{fig:pretraining-val-loss} shows that our proposed methods lead to faster convergence and further stabilizes the pre-training.

\begin{table}[htbp]
\caption{Baseline and our proposed optimized setup, pre-trained and fine-tuned on the whole 960 hours of LibriSpeech.}
\begin{center}
\begin{tabular}{|c|c|c|c|c|c|}
\hline
 & \multicolumn{4}{|c|}{\textbf{WER}} \\
\hline
 & \textbf{dev-clean} & \textbf{dev-other} & \textbf{test-clean} & \textbf{test-other} \\
\hline
\underline{baseline} & \underline{5.69}   & \underline{15.91}   & \underline{5.93} & \underline{16.58} \\
ours & 4.33 & 11.92 & 4.52 & 11.50 \\
\hline
\end{tabular}
\label{tab:full-results}
\end{center}
\end{table}

\section{Discussion}
We proposed improvements to the BEST-RQ training method for self-supervised learning in speech recognition.
These modifications include the implementation of a multi-codebook setup, the incorporation of an auxiliary Kullback-Leibler divergence loss for regularization, and the introduction of feature cluster-specific codebooks derived from low-level audio features. 
Together, these improvements offer enhancements in both performance and training stability.

The use of multiple codebooks enhances the system's performance by an 11.2\% relative improvement in WER compared to the baseline setup in preliminary experiments.

The combination of CE loss and KL-divergence as a regularizing loss further improves performance, with a 7.8\% reduction in WER in comparison to using only CE loss.
The KL-divergence helps the model maintain stable predictions across codebooks by reducing divergence between the predicted and target distributions.

Introducing feature-level clustering of utterances, with codebooks dedicated to specific clusters, offers further improvements. 
This clustering ensures that codebooks are specialized for particular types of audio input derived from low-level features.

Across all experiments, the proposed modifications show an ability to achieve faster convergence, as observed in the validation loss.

\section{Limitations}
The weighting factor of the KL-divergence loss is a hyperparameter, which heavily affects the regularizing effect. Extreme weighting factors result in decreased WER and have to be selected individually for the data used.

The multi-codebook approach increases the model’s complexity and size in pre-training. However, in fine-tuning, this overhead is omitted, since only the conformer encoder is used for downstream tasks, resulting in the same model size.

While the proposed modifications result in improvements in preliminary experiments, the effect of each modification on the full pre-training and fine-tuning still needs to be explored.

\section{Conclusion}
We proposed optimizations to the BEST-RQ setup and demonstrate notable improvements in both WER (up to 30.6\%) and training stability for speech recognition tasks.
The multi-codebook approach, auxiliary loss integration, and cluster-specific codebooks lead to a more robust and generalizable model. 
The results on LibriSpeech show that the proposed modifications effectively advance self-supervised learning using BEST-RQ in speech recognition.

Future work could explore refining the weighting of the KL-divergence loss and investigating further extensions of the multi-codebook strategy to additional audio tasks. 
Moreover, applying these methods to larger and more diverse speech datasets may provide further insight into their scalability and adaptability.


\newpage

\bibliographystyle{IEEEtran}
\bibliography{IEEEabrv,IEEEexample}

\end{document}